\documentclass[twocolumn]{jpsj2} %% two-column layout
\title{AC/DC Susceptibility of the Heavy-Fermion Superconductor CePt$_3$Si under Pressure}

\author{%
Yoshihiro \textsc{Aoki}$^{1}$, Akihiko \textsc{Sumiyama}$^{1}$\thanks{E-mail address: sumiyama@sci.u-hyogo.ac.jp}, Gaku \textsc{Motoyama}$^{1}$, Yasukage \textsc{Oda}$^{1}$, Takashi \textsc{Yasuda}$^{2}$, Rikio \textsc{Settai}$^{2}$ and Yoshichika \textsc{\={O}nuki}$^{2}$}

\inst{%
$^1$Department of Material Science, Faculty of Science, University of Hyogo, Ak\={o}-gun 678-1297\\
$^2$Graduate School of Science, Osaka University, Toyonaka
560-0043\\
}

\recdate{\today}

\abst{

We have investigated the pressure dependence of ac and dc susceptibilities ($\chi_{\rm{ac}}$ and $\chi_{\rm{dc}}$) of the heavy-fermion superconductor CePt$_3$Si ($T_{{\rm c}}$= 0.75 K) that coexists with antiferromagnetism ($T_{{\rm N}}$= 2.2 K). As hydrostatic pressure is increased, $T_{{\rm c}}$ first decreases rapidly, then rather slowly near the critical pressure $P_{{\rm c}}$ = 0.6 GPa and shows a stronger decrease again at higher pressures, where $P_{{\rm c}}$ is the pressure at which $T_{{\rm N}}$ becomes zero. A transition width and a difference in the two transition temperatures defined in the form of structures in the out-of-phase component of $\chi_{\rm{ac}}$ also become small near $P_{{\rm c}}$, indicating that a double transition observed in CePt$_3$Si is caused by some inhomogeneous property in the sample that leads to a spatial variation of local pressure. A sudden increase in the Meissner fraction above $P_{{\rm c}}$ suggests the influence of antiferromagnetism on superconductivity. 

}
\23
\kword{magnetic susceptibility, pressure, CePt$_{3}$Si, heavy-fermion superconductor, Meissner effect}

\begin{document}
\maketitle

\newpage
\section{Introduction} %% No sections necessary for express letters, letters and short notes
Ever since the discovery of heavy-fermion superconductivity in CePt$_3$Si,\cite{1} many studies have been made to clarify its unconventional superconductivity. The temperature dependence of various physical properties, such as thermal conductivity\cite{2} and magnetic penetration depth,\cite{3} suggests a gap function vanishing on the Fermi surface. The large upper critical field\cite{1} and NMR Knight shift\cite{4} suggest its spin triplet superconductivity, while the nuclear spin-lattice relaxation rate has a coherence Hebel-Slichter peak at $T_{{\rm c}}$,\cite{5} indicating conventional superconductivity. The possibility of a mixed spin singlet and triplet state that is allowed owing to the lack of the center of inversion symmetry in the crystal structure has been proposed to explain these results.\cite{6,7}

Besides unconventional superconductivity, CePt$_{3}$Si possesses a complex pressure-temperature ({\it P}-{\it T}) phase diagram. Antiferromagnetism ($T_{{\rm N}}$ = 2.2 K) and superconductivity ($T_{{\rm c}}$ = 0.75 K) coexists at ambient pressure,\cite{1} and both $T_{{\rm N}}$ and $T_{{\rm c}}$ decrease with an increase in pressure.\cite{8} Antiferromagnetism disappears above $P_{{\rm c}}$ = 0.6 GPa, while superconductivity is observed up to 1.5 GPa. In addition, the superconducting transition always shows a large transition width at $\sim 0.2$ K,\cite{9,10} and some samples even show a double peak in specific heat measurements.\cite{11,12,13} Although the possibility that the second transition is due to a trace of antiferromagnetic phase\cite{12} or unconventional superconductivity\cite{11} has been proposed, our recent investigation suggests that CePt$_{3}$Si contains two superconducting phases with different $T_{{\rm c}}$'s.\cite{13} 

In this work, we investigate the pressure dependence of the double superconducting transition by magnetic susceptibility measurements and elucidate what causes the difference in $T_{{\rm c}}$ inside CePt$_{3}$Si. We also report the change in the Meissner effect below and above $P_{{\rm c}}$.  
\section{Experimental}

We studied the single crystal of CePt$_3$Si denoted as C-1 in our previous paper,\cite{13} which shows zero resistivity at 0.63 K and a single transition at 0.4 K in specific heat measurements. Details of the producing process are provided in previous papers.\cite{9,10} For comparison, a polycrystalline sample, which has a nominal composition of Ce$_{1.01}$Pt$_3$Si and was prepared by arc melting method, was used also.\cite{14} Both samples have a rectangular shape with edges of the order of mm. A magnetic field was applied along the [100] ($a$ axis) direction for the single crystal.

The sample was mounted in a small mutual inductance coil. A piece of superconducting In was  mounted also to determine pressure by the shift of $T_{{\rm c}}$. For the application of pressure, the coil was set in a piston cylinder cell with Daphne oil (7373) as pressure medium. Both $\chi_{\rm{ac}}$ and $\chi_{\rm{dc}}$ were measured using a SQUID and the coil mentioned above. Details of the measuring technique at low temperatures have been described in our previous paper.\cite{13} 
%%%%%%%%%%%%%%%%%%%%%%%%%%%%%%%%%%%%%%%%%%%%%%%%%%%%%%%%%%%%%%%%%%%%%%%
%Fig.1 
% 
%
\begin{figure}
\begin{center}
\includegraphics*[width=0.6\linewidth, trim=1cm 3cm 2cm 5cm, clip]{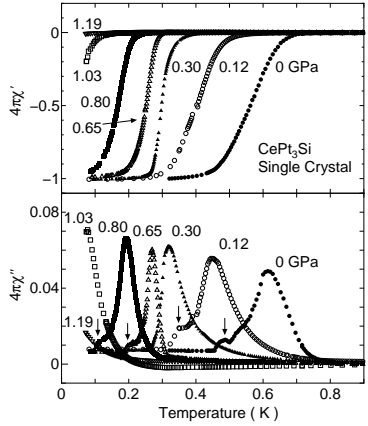}
\caption{\label{fig1}Temperature dependence of ac magnetic susceptibility ($\chi_{{\rm ac}}$ = $\chi '$ + $i \chi ''$) of a CePt$_{3}$Si crystal under different pressures. $\chi_{{\rm ac}}$ was measured with an ac field of $H_{{\rm ac}}$ = 1.2 mOe at 160 Hz. The arrows indicate additional shoulder-like structures in $\chi ''$.}
\end{center}
\end{figure}
%
%%%%%%%%%%%%%%%%%%%%%%%%%%%%%%%%%%%%%%%%%%%%%%%%%%%%%%%%%%%%%%%%%%%%%%%

\section{Results and Discussion}
Figure \ref{fig1} shows the temperature dependence of ac susceptibility $\chi_{{\rm ac}}$ = $\chi '$ + $i \chi ''$ for the single crystal under hydrostatic pressure, where $\chi '$ is the in-phase component to the current in the primary coil and $\chi ''$ is the out-of-phase or energy dissipation component. Since full diamagnetism at ambient pressure was observed in the previous work,\cite{13} the values of $\chi '$ at the highest temperature are set to 0 and the value at each temperature is calculated by comparison with full diamagnetism ($4\pi\chi '= -1$) at $P$ = 0 GPa. In addition to the decrease in $T_{{\rm c}}$ that is already reported,\cite{9,8,15} it is found that the transition width becomes small between $P$= 0.3 and 0.65 GPa. The second transition that is characterized by a shoulder-like structure in $\chi ''$ also becomes obscure at $P$= 0.3 GPa because of the sharp transition.  It should be noted that this transition is not the recently proposed second transition that is from the $d_{xz}$-wave to the $d_{xz}\pm id_{yz}$-wave state, since the latter transition should be observed only below $P_{{\rm c}}$.\cite{16}

\begin{figure}
\begin{center}
\includegraphics*[width=0.6\linewidth, trim=1cm 8cm 2cm 8cm, clip]{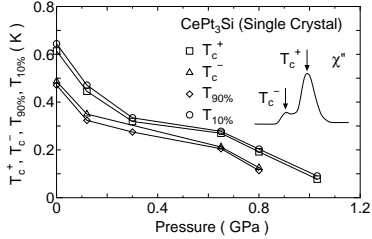}
\caption{\label{fig2}Pressure dependence of four characteristic temperatures: $T_{{\rm c}}^{+}$, $T_{{\rm c}}^{-}$, $T_{{\rm 90\%}}$, and $T_{{\rm 10\%}}$. $T_{{\rm 90\%}}$ and $T_{{\rm 10\%}}$ are the temperatures where the CePt$_{3}$Si crystal shows 90\% and 10\% of full diamagnetism, respectively. As indicated schematically in the inset, $T_{{\rm c}}^{+}$ and $T_{{\rm c}}^{-}$ are the temperatures where $\chi ''$ reaches a maximum and has a local maximum or a shoulder, respectively. The solid lines through the data points are guides to the eye.}
\end{center}
\end{figure}

In order to demonstrate the pressure dependence of superconducting transition more clearly, we define four characteristic temperatures $T_{90\%}$, $T_{10\%}$, $T_{{\rm c}}^{+}$, and $T_{{\rm c}}^{-}$ as follows: $\chi ''$ has a maximum at $T_{{\rm c}}^{+}$ and below $T_{{\rm c}}^{+}$ it shows a local maximum or a shoulder at $T_{{\rm c}}^{-}$, while $T_{90\%}$ and $T_{10\%}$ are the temperatures where $\chi '$ is 90 \% and 10 \% of full diamagnetism, respectively. The results are shown in Fig. \ref{fig2}. As the pressure is increased, the four temperatures first decrease steeply up to $P$ = 0.3 GPa, then change slowly, and begin a stronger decrease above $P$ = 0.65 GPa. A similar behavior has already been reported in specific heat measurements of a single crystal.\cite{8} The slow decrease in $T_{{\rm c}}$ probably reflects the vanishing of antiferromagnetism at $P_{{\rm c}}$ = 0.6 GPa.

%%%%%%%%%%%%%%%%%%%%%%%%%%%%%%%%%%%%%%%%%%%%%%%%%%%%%%%%%%%%%%%%%%%%%%%
%Fig.2
% 
%
%
%%%%%%%%%%%%%%%%%%%%%%%%%%%%%%%%%%%%%%%%%%%%%%%%%%%%%%%%%%%%%%%%%%%%%%%
In our previous paper,\cite{13} we have reported AC/DC susceptibility of the present crystal and a crystal that shows two peaks at $T_{{\rm c}}^{+}$ and $T_{{\rm c}}^{-}$ in the specific heat; diamagnetic susceptibility and the Meissner effect of the latter sample is larger than that of the present sample near $T_{{\rm c}}^{+}$, and yet no anomalies is observed at $T_{{\rm c}}^{-}$. This result suggests that the double peaks in the specific heat is ascribed to the coexistence of superconducting phases with different $T_{{\rm c}}$'s rather than the successive change of a superconducting phase into another state at $T_{{\rm c}}^{-}$; the crystal with the double peaks contains a greater amount of the superconducting phase with $T_{{\rm c}}^{+}$ and a smaller amount of the the superconducting phase with $T_{{\rm c}}^{-}$ in comparison with the present crystal that shows a single peak at $T_{{\rm c}}^{-}$ in the specific heat. However, the origin of the difference in $T_{{\rm c}}$ was still unclear.

In Fig. \ref{fig2}, the difference between $T_{{\rm c}}^{+}$ and $T_{{\rm c}}^{-}$ as well as the transition width $\Delta T=T_{90\%}-T_{10\%}$ becomes small at pressures between 0.30 GPa and 0.65 GPa where $T_{{\rm c}}$ depends little on pressure. This result suggests that some inhomogeneous property, such as the strain field around crystallographic defects and a deviation from the stoichiometric composition, causes a variation in local pressure and the resultant $T_{{\rm c}}$ difference according to the $P-T_{{\rm c}}$ characteristics. Even if such a pressure variation exists, the difference in local $T_{{\rm c}}$ becomes small and the transition becomes sharp, when the applied pressure is near $P_{{\rm c}}$ and $T_{{\rm c}}$ depends little on pressure. Since $T_{{\rm c}}$ determined by specific heat measurements corresponds to $T_{{\rm c}}^{-}$ in this single crystal, a large part of the sample has $T_{{\rm c}}^{-}$ and a small part has $T_{{\rm c}}^{+}$ probably because of lower local pressures. One possible explanation for such a local reduction in pressure is a negative pressure caused by lattice defects or partial replacement of elements.

The result that an increase in pressure reduces the transition width $\Delta T=T_{90\%}-T_{10\%}$ at least below 0.3 GPa contrasts with what is observed in resistivity measurements; the onset temperature changes little, and the transition becomes broader\cite{15} or shows a complex behavior\cite{9} with increasing pressure. Since only a small amount of superconducting phase can cause a drop in resistivity, a trace of low pressure region ($P\sim 0$ GPa) that is only detectable in resistivity measurements may still remain under pressure. 

%%%%%%%%%%%%%%%%%%%%%%%%%%%%%%%%%%%%%%%%%%%%%%%%%%%%%%%%%%%%%%%%%%%%%%%
%Fig.3 
% 
%
\begin{figure}
\begin{center}
\includegraphics*[width=0.6\linewidth, trim=1cm 3cm 2cm 5cm, clip]{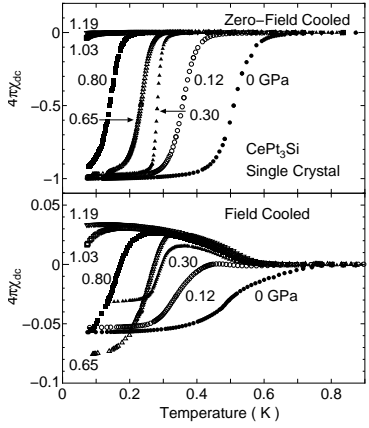}
\caption{\label{fig3}Temperature dependence of $\chi_{{\rm dc}}$ under different pressures for the CePt$_{3}$Si crystal, as obtained by the field-cooled and the zero-field cooled methods. The applied fields is 0.12 Oe. }
\end{center}
\end{figure}
%
%%%%%%%%%%%%%%%%%%%%%%%%%%%%%%%%%%%%%%%%%%%%%%%%%%%%%%%%%%%%%%%%%%%%%%%

Although ac diamagnetic susceptibility reflects bulk superconductivity in comparison with zero resistivity in the sense that the shielding current on the whole surface is detected, dc susceptibility measurements that detect the Meissner effect gives more information about bulk properties, as shown in Fig. \ref{fig3}. The measuring procedure is as follows. The sample was first cooled down to the lowest temperature in zero magnetic field and then $H_{{\rm dc}}$ was applied. The diamagnetic susceptibility due to the shielding current was measured during the warming process up to temperatures well above $T_{{\rm c}}$ (ZFC: zero-field cooled). In the same field, the Meissner effect was measured during the cooling process (FC: field cooled).

Apart from a slight decrease in $T_{{\rm c}}$ due to a larger applied field, the ZFC susceptibility shows a behavior similar to the ac susceptibility $\chi '$ in Fig. \ref{fig1}. The FC susceptibility, on the other hand, shows a complicated behavior. At ambient pressure, the increasing rate of the Meissner effect with decreasing temperature slightly changes at about 0.5 K, indicating the second transition. When pressure is applied, the second transition is difficult to observe in FC susceptibility measurements probably because the superconducting transition becomes sharp. In addition, a paramagnetic signal appears below 0.6 K at pressures above 0.3 GPa, which is canceled out by a diamagnetic (Meissner) signal below $T_{{\rm c}}$. The magnitude and the temperature where the paramagnetic signal appears depends little on pressure.
%%%%%%%%%%%%%%%%%%%%%%%%%%%%%%%%%%%%%%%%%%%%%%%%%%%%%%%%%%%%%%%%%%%%%%%
%Fig.4 
% 
%
\begin{figure}
\begin{center}
\includegraphics*[width=0.6\linewidth, trim=1cm 3cm 2cm 5cm, clip]{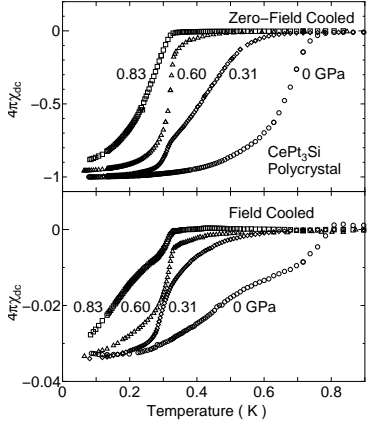}
\caption{\label{fig4}Temperature dependence of $\chi_{{\rm dc}}$ under different pressures for a CePt$_{3}$Si polycrystal, as obtained by the field-cooled and the zero-field cooled methods. The applied field is 0.12 Oe.}
\end{center}
\end{figure}
%
%%%%%%%%%%%%%%%%%%%%%%%%%%%%%%%%%%%%%%%%%%%%%%%%%%%%%%%%%%%%%%%%%%%%%%%

In order to test whether the observed paramagnetic anomaly is ascribed to bulk CePt$_{3}$Si or some impurity phase, we have measured dc susceptibility of a polycrystalline sample, as shown in Fig. \ref{fig4}. The paramagnetic signal that appears under high pressures in Fig. 3 has not been observed for the polycrystalline sample, indicating that it is a sample-dependent property. The paramagnetic signal for the single crystal is remarkable only in the FC susceptibility and resembles the ferromagnetic anomaly that is observed at 3 K for samples with small variations in the composition of Ce$_{1+x}$Pt$_{3+y}$Si$_{1+z}$.\cite{14} Considering also the complexity of the Ce-Pt-Si ternary phase diagram,\cite{17} a trace of some second phase may cause the paramagnetic anomaly observed only for the single crystal. Other differences between the two samples will be discussed later.
%%%%%%%%%%%%%%%%%%%%%%%%%%%%%%%%%%%%%%%%%%%%%%%%%%%%%%%%%%%%%%%%%%%%%%%
%Fig.5
% 
%
\begin{figure}
\begin{center}
\includegraphics*[width=0.7\linewidth, trim=0cm 10cm 2cm 8cm, clip]{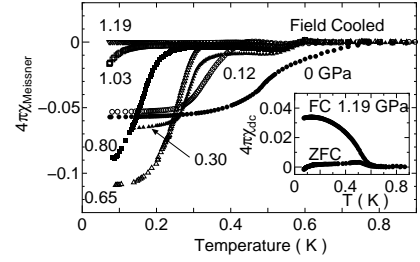}
\caption{\label{fig5}Temperature dependence of $\chi _{{\rm Meissner}}$ for the CePt$_{3}$Si crystal, where $\chi _{{\rm Meissner}}$ values at $P$ = 0 and 0.12 GPa are the same as FC (Field Cooled) $\chi_{{\rm dc}}$ in Fig. \ref{fig3}. At $P\ge$ 0.30 GPa, FC $\chi_{{\rm dc}}$ at $P$ = 1.19 GPa, which is shown in the inset, is subtracted from FC $\chi_{{\rm dc}}$ below 0.6 K in order to remove the paramagnetic contribution. }
\end{center}
\end{figure}
%
%%%%%%%%%%%%%%%%%%%%%%%%%%%%%%%%%%%%%%%%%%%%%%%%%%%%%%%%%%%%%%%%%%%%%%%

In the field-cooling process in Fig. \ref{fig3}, the paramagnetic signal that appears above  0.30 GPa shows almost the same temperature dependence until the superconducting transition starts. Since at $P$= 1.19 GPa only a slight change due to superconductivity is observed below 0.12 K, we regard FC $\chi _{{\rm dc}}$ at $P$= 1.19 GPa below 0.6K as the underlying paramagnetic contribution and subtract it from FC $\chi _{{\rm dc}}$ at $P\ge  0.30$ GPa to extract the Meissner effect. The result is shown in Fig. \ref{fig5}. It is obvious that the paramagnetic contribution is overestimated for $\chi _{{\rm dc}}$ at $P$= 0.30 GPa and the Meissner fraction $-4\pi \chi _{{\rm Meissner}}$ should be a little smaller. The result that the Meissner fraction at the lowest temperatures is almost the same between $P$=0.12 and 0.30 GPa suggests that the emergence of the paramagnetic signal does not change the magnitude of the Meissner effect. In contrast, the abrupt increase in the Meissner effect between $P$=0.30 and 0.65 GPa is clearly seen, indicating that it originates from the disappearance of antiferromagnetism at $P_{{\rm c}}$= 0.6 GPa.

Even if the whole volume is superconducting, the Meissner fraction in real samples of type-II superconductors is usually much smaller than 1 because of vortex pinning. The result that the Meissner fraction at ambient pressure has not changed before and after the application of pressure suggests that the abrupt increase above $P_{{\rm c}}$ can not be attributed to pressure-induced irreversible change of the lattice that acts as a pinning site. Since vortex pinning occurs by the local variation of the free energy of a flux line that depends on coherence length $\xi$, penetration depth $\lambda$ and critical field $H_{{\rm c}}$, the disappearance of antiferromagnetism may cause an abrupt change in these superconducting properties. One possible explanation for such a change is that some of the nodes in the energy gap of CePt$_{3}$Si vanish when antiferromagnetism disappears.\cite{16}

In Fig. \ref{fig4}, the polycrystalline CePt$_{3}$Si under pressure shows a less decrease in the onset temperature and consequently a broader transition width than the single crystal, suggesting that the polycrystalline sample is less homogeneous with respect to local pressure than the single crystal. A partial transition at 0.3 K may be explained by the $P-T$ characteristics in Fig. 2; the pressure range in which $T_{{\rm c}}$ is close to 0.3 K is relatively wide, so that many parts tend to have local pressure values in this range and the transition at 0.3 K occurs. This may also explain the drop of the onset temperature from 0.8 K to 0.3 K observed in resistivity measurements under pressure.\cite{9}

Since the grain boundaries in a polycrystal act as additional pinning sites, the Meissner fraction tends to be smaller than that of a single crystal, as seen in Figs. 4 and 5. They may also prevent the increase in the Meissner fraction above $P_{{\rm c}}$; even if more magnetic fluxes are expelled from the crystal grains by the absence of antiferromagnetic order, the grain boundaries probably trap at least a part of them. The comparison of FC $\chi _{{\rm dc}}$ well above $P_{{\rm c}}$ indicates that the Meissner fraction at $P$ = 0.80 GPa is larger than that below $P_{{\rm c}}$ at the lowest temperatures for the single crystal, while the Meissner fraction at $P$ = 0.83 GPa is smaller than that below $P_{{\rm c}}$ for the polycrystal, although both samples show about 90 \% of full diamagnetism  in ZFC $\chi _{{\rm dc}}$. Moreover. the onset of superconductivity in ZFC $\chi _{{\rm dc}}$ is 0.3 K for the polycrystal, indicating a certain amount of the superconducting phase may still coexist with antiferromagnetism. The inhomogeneous distribution of local pressure together with the vortex pinning at the grain boundaries in the polycrystal makes it difficult to observe the anomalies near $P_{{\rm c}}$ that appeared for the single crystal: the decrease in the transition width and the change in the Meissner fraction.

\section{Conclution}
In conclusion, ac and dc magnetic susceptibility measurements of CePt$_{3}$Si under pressure have revealed that the decrease in the slope of the pressure dependence of $T_{{\rm c}}$ and the decrease in the superconducting transition width occur at about the critical pressure $P_{{\rm c}}$ where coexisting antiferromagnetism vanishes. The large transition width and the sample-dependent second transition at ambient pressure are well explained on the assumption that a spatial variation of local pressure exists in CePt$_{3}$Si. Another anomaly near $P_{{\rm c}}$, that is, the abrupt increase in the Meissner fraction above $P_{{\rm c}}$ provides some information on the relation between superconductivity and antiferromagnetism.

%
%Acknowledgement

\section*{Acknowledgments}
This study was partly supported by a Grant-in-Aid from the Ministry of Education, Culture, Sports, Science and Technology (MEXT), Japan. One of us (Y. \={O}.) was financially supported by the Grant-in-Aid for COE Research (10CE2004) of the MEXT, Japan.

\end{document}